\documentclass{llncs}

\usepackage{t1enc}
\usepackage[cp1250]{inputenc}

\usepackage{latexsym}
\usepackage{amssymb}
\usepackage{tabularx}
\usepackage{amsfonts}
\usepackage{listings} 

\usepackage{colortbl}
\usepackage{xcolor}
\usepackage{pst-all} 
\usepackage{pst-poly}
\newpsobject{showgrid}{psgrid}{subgriddiv=1,griddots=10,gridlabels=6pt}
\usepackage{graphicx}
\usepackage{fancybox}
\usepackage{wrapfig}

\usepackage{algorithm}
\usepackage{algpseudocode}

\newcommand{\con}{\wedge} 
\newcommand{\dis}{\vee} 
\newcommand{\alw}{\Box} 
\newcommand{\imp}{\Rightarrow} 
\newcommand{\equ}{\Leftrightarrow} 
\newcommand{\som}{\Diamond} 

\newcommand\nonofoot[1]{%
   \begingroup
   \renewcommand\thefootnote{}\footnote{\kern-0.2ex#1}%
   \addtocounter{footnote}{-1}%
   \endgroup
}

\usepackage{doc}
\usepackage{makeidx}

\begin{document}

%
\title{Context-awareness of the IoT through the on-the-fly preference modeling}



%
\author{Rados{\l}aw Klimek \and Leszek Kotulski}

\institute{AGH University of Science and Technology,\\
           al.\ A.\ Mickiewicza 30, 30-059 Krakow, Poland\\
           \email{\{rklimek,kotulski\}@agh.edu.pl}
           }



\clearpage

\maketitle

\begin{abstract}
The context-awareness of things that belong to IoT networks have to
be considered in a distributed computation paradigm.
In the paper we suggest the use of graph transformations and temporal logic as a formal framework for
a knowledge representation of user/inhabitant behaviors in multi-agent systems.
IoT networks are considered as graph structures.
Dynamic preference models, understood as a priority in the selecting, is also introduced.
Preference models as a result of observed behaviors base on formal logic,
and they are built on-the-fly by software agents.
Software agents gather knowledge about user preferences expressed in terms of logical specifications
as well as suggest on-the-fly future behavior basing on the logical inference process using the semantic tableaux method.
The predictive processes are result of some new and important events in the context of IoT systems that should meet a response.
Due to the ubiquitous availability of cyber systems that interact with physical environments,
there is a great need to develop technologies that target the whole IoT system as a context-awareness system.
Formal approach increases the trustworthy of a system.
A simple yet illustrative example is provided.
\end{abstract}

\setcounter{tocdepth}{2}

%
%

\pagestyle{empty}

\section{Introduction}
\label{sec:introduction}

\thispagestyle{empty}

The Internet of Things, or IoT, refers to uniquely identifiable objects
enable automatic transfer data over a network and cooperation without any kind of intervention.
There are some examples of the IoT successful implementations,
e.g.\ smart bus stops and smart parking spots both in Barcelona (Spain),
c.f.\ \cite{Kamel-Boulos-Al-Shorbaji-2014}.
They not only improved the quality of live but also,
in the case of parking spots,
increased revenues of parking fees as well as allow to create new jobs.
Success attracts success.
The UK government is going to spend huge sums of money for the IoT technology~\cite{BBC-2014}.
IoT becomes increasingly ubiquitous and revolutionizes pervasive computing and its applications.

Pervasive computing is understood as existing or being everywhere at the same time,
assuming the omni-presence of computing providing strong support for users/inhabitants.
Ubiquitous computing is a post-desktop information processing organizing
an environment in always unobtrusive and available devices of any type,
and communicating through easy available and effective networks of any kind.
It means being and seeming to be everywhere spreading the computation into everything
around us/user/inhabitant.
Thus, computing is embedded everywhere in the environment providing freely available computations.
Computing machines must sense user/inhabitant presence and act accordingly.
This leads to the need for awareness of computational processes and applications.
Context-awareness is a property of linking changes in the environment with computer systems
which are otherwise static.
Important aspects of context are: where you are, who you are with,
and what resources are nearby~\cite{Dey-Abowd-2000}.
Context-awareness is necessary to anticipate user/inhabitant's needs and pro-actively offer functions and services.
In software engineering, for software developers,
context-awareness means sensing and reacting on environment,
where remote sensors, devices and monitoring technology might
allow the continuous capture and analysis of the incoming data.
This constitutes significant challenge for developers.
IoT should be environmentally safe,
what is especially important in the case of safety-critical systems or functions,
thus the reliability demand for such systems increases.
Deployment of a formal approach during the development phase
might be an appropriate response to a such demand.
Sophisticated sensors and algorithms require a special attention.
Solid and reliable interactions of IoT systems entail technologies basing on a formal approach.

In this paper we suggest the use of formal logic and graph transformations which are formal background
for a multi-agent system which operates in the IoT environment to provide
context-aware and pro-active services.
To the best of our knowledge,
there is a lack of such deployment for both formalisms in the mentioned above domains/applications.
The main issue addressed in the paper is to prepare a formal foundation of
the hierarchical multi-agent structure supporting the IoT system that is verified by
finding a workable and comprehensive solution and analyzing a simple yet illustrative example
that discerns intent rather than mere inputs.

Broadly speaking, the proposed multi-agent system operates in such a way that
gathers knowledge about users/inhabitants' behaviors and preferences and
guides further and possible comfortable behaviors in the IoT network which is considered as a graph structure.
Dynamic preference models, understood as a priority in the selecting, is also introduced.
Preference models as a result of observed behaviors base on formal logic,
and they are built on-the-fly by software agents.
Software agents gather knowledge about user preferences expressed in terms of logical specifications
as well as suggest on-the-fly future behavior basing on the logical inference process using the semantic tableaux method.

The IoT environment should be formally modeled in order to allow
the computational agents make some decisions.
Graph seems to be intuitive formalism to represent the static structure of the IoT networks.
Graph transformations model this network modification.
Unfortunately, in most cases we have a problem with the synchronization of the parallel work of a few agents on the same structure.
In this paper we suggest using the replicated complementary graph RCG concept~\cite{Kotulski-Sedziwy-2011-icannga}
that allows to split the graph into set of complementary subgraphs
that represent part of knowledge maintained by an agent.
The RCG concept introduce the explicit synchronization mechanism that guarantees correct application of
the transformation rules designed for the centralized system in the distributed multi-agents environments.

Preferences understood as a priority in the selecting of something over another or
others follows from observed users/inhabitant's behaviours.
However, preference models is built in a different way comparing
works~\cite{Klimek-2013-peccs,Klimek-Wojnicki-Ernst-2013-icaisc,Klimek-2013-icmmi},
where the introduced approach is static, i.e.\ not changed during the execution of a system.
The goal of this approach is building preference models on-the-fly when the system operates.
Preference models base on temporal logic,
and they are built by software agents sensing and reacting users/inhabitants
and preparing user-oriented preference decisions.
Formal logic allows to register behavior in a precise way,
i.e.\ without any ambiguity typical for natural languages.
``Logic has simple, unambiguous syntax and semantics.
It is thus ideally suited to the task of specifying information systems''~\cite{Chomicki-Saake-1998}.
Moreover, it also allows to perform automatic reasoning to obtain preference decisions for newly observed users/inhabitants.
The semantic tableaux method for temporal logic, as a method of reasoning, is proposed.
Logical inference allows to build truth trees, 
searching for satisfiability or contradictions with respect to logical specifications.

The parking space system is provided as an example.
It is considered and understood as a graph structure, where every thing/object is a graph node.
The past behavior of users/inhabtants are registered
in the form of logical specifications expressed in terms of temporal logic formulas.
There is a hierarchy of software agents in the system.
They gather basic information in nodes,
identify users/inhabitants,
observe their behaviours,
and register behaviours in the form of logical specifications,
preparing preference decisions, as a suggested behaviours, for users/inhabitants.
Some agents exist permanently and some of them exist temporarily.
The whole system constitutes a smart, context-aware and pro-active software system.
Some discussion that refers related works is in the next Section.

Let us focus our attention on some related works.
In paper by Perera et al.~\cite{Perera-etal-2014} a wide range of problems which concern the IoT is discussed.
An analysis and research that allow to understand challenges of context-aware computing are presented.
A set of projects is evaluated, and some trends are discussed.
In paper by Gubbi et al.~\cite{Gubbi-etal-2013} a proliferation of devices is in focus,
and also a cloud based model for the IoT emerging technology is discussed.
In paper by Ye et al.~\cite{Ye-etaal-2012} situation identifications for data which come from
a huge number of sensor are discussed.
There are considered many different situation and specification techniques.
Work by Atzori et al.~\cite{Atzori-etal-2010} contains a detailed report on different aspects on the IoT.
As pointed out,
the US National Intelligence Council included the IoT in the list of
six ``Disruptive Civil Technologies'' with potential impacts on US national power~\cite{NIC-2008}.
All these issues highlight the importance of IoT for everyday life.
This paper is a continuation of the work~\cite{Klimek-Kotulski-2014-IE-AITAmI}.

Summing up,
it seems that the proposed approach basing on this formal framework
provides, hopes promising, improved methods to monitor the IoT environment as a smart system,
i.e.\ both sensing and reacting the environment.
It allows continuous realtime data collection and reliable analysis via remote devices
enriching user/inhabitan's everyday situations.

\section{Context models and preferences}
\label{sec:context-model}

Issues to the Internet of Things, context-awareness systems,
preferences and their models as well temporal logic are discussed in the Section.
Pervasive computing or ubiquitous computing can be understood
as existing or being everywhere at the same time,
assuming the omni-presence of computing providing strong
support for users/inhabitants.
Because of the pervasiveness of everyday technologies,
one use them without thinking about them,
making the technology effectively invisible to the user.
Context-awareness and context modeling is one of the crucial aspect
of pervasive systems.
The Internet of Things (IoT) is another important subject that refers to
pervasive computing and context-aware systems.
It could be understood as a scenario in which objects,
users, inhabitants (or even animals) are provided with
unique identifiers to enable automatic transfer data over
a network and cooperation without any kind of intervention.
The aim is to give strong support for users or inhabitants.

\begin{figure}[htb]
\centering
\begin{pspicture}(2.0,3.2) 
\rput(1.4,0.4){\rnode{s1}{}}
\rput(1.4,1.8){\rnode{s2}{}}
\ncline[linewidth=1pt]{->}{s1}{s2}\naput[labelsep=-13pt,nrot=:D]{\textsf{tracking}}
\rput(1.4,1.9){\rnode{s3}{}}
\rput(1.4,3.1){\rnode{s4}{}}
\ncline[linewidth=1pt]{->}{s3}{s4}\naput[labelsep=-13pt,nrot=:D]{\textsf{sensing}}
\rput(.6,3.1){\rnode{s5}{}}
\rput(.6,1.9){\rnode{s6}{}}
\ncline[linewidth=1pt]{->}{s5}{s6}\naput[labelsep=-13pt,nrot=:D]{\textsf{reacting}}
\rput(.6,1.8){\rnode{s5}{}}
\rput(.6,0.4){\rnode{s6}{}}
\ncline[linewidth=1pt]{->}{s5}{s6}\naput[labelsep=-13pt,nrot=:D]{\textsf{influencing}}
\end{pspicture}
\includegraphics[width=.55\textwidth]{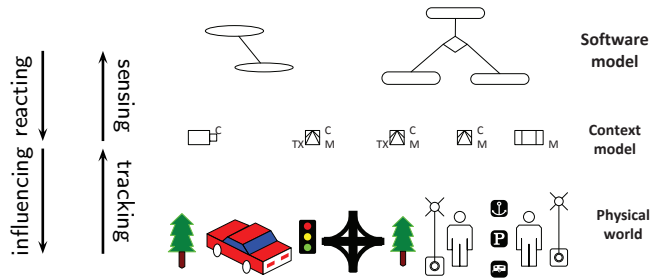}
\caption{A three-layer context model of an smart environment}
\label{fig:context-aware-systems}
\end{figure}

Context-awareness is an important component of today-s most
pervasive applications which behaviour is characterized by
the interpretation logic that is embedded inside these applications.
This type of computing assumes transfer of contextual information among
pervasive applications in the IoT network.
A context is conditions and circumstances that are relevant to
the working system.
A sample physical world which create a context which is interpreted by
context-aware applications is shown in Fig.~\ref{fig:context-aware-systems},
c.f.\ also~\cite{Bettini-etal-2010}.
The physical world and the context-awareness software constitute
the smart environment.
Context model creates different types of sensors
which are distributed in the whole considered physical area.
For example, the smart street lighting could be controlled by intelligent software
which is context-awareness and pro-active.
Distributed sensors constitutes a kind of eyes for software systems.
These ideas also refer to the concept of \emph{Ambient Intelligence} (AmI),
i.e.\ electronic devices that are sensitive and responsive to the presence of
humans/inhabitants.
It follows that the smart application must
both understand the context (tracking and sensing),
that is context-aware, and characterized by pro-activity,
that is acting in advance to deal with an expected occurrences or situations,
especially negative or difficult ones (reacting and influencing).
Context-aware system is able to adapt their operations to
the current context without explicit user intervention.
It follows that it requires special treatment when modeling software.
Multi-agent systems seem to be a proper idea to support the issues of this work.


Issues to preferences and their models as well temporal logic and deductive reasoning are
discussed in the Section.
Preference modeling is a key step in many fields. It enables
customization of software behavior to users needs.
The construction of preference models is particularly important in
systems of pervasive computing.
Preference modeling needs formalization and it is discussed in
some works, e.g.~\cite{Ozturk-etal-2005}. The model of preference
might be constructed using fuzzy sets, classical logic and
many-valued logics. Classical logic, and particulary rule-based
systems, are especially popular~\cite{Fong-etal-2011}.
Non-classical logics, and especially temporal logic, are less
popular. On the other hand, temporal logic is a well established
formalism for describing reactiveness, and meanwhile, the typical
pervasive application should be characterized by reactivity and
flexibility in adapting to changes on the user side. These changes
may result from recognized and predefined preferences. The
variability and change in valuation of logical statements are
difficult to achieve in classical logic. ``Certainly there are
unchanging truths, but there are changing truths also, and it is a
pity if logic ignores
these''~(A.~Prior\footnote{Arthur N.\
Prior's the significant achievement was the invention of tense
logic with new modal operators. (He was a logician entirely
educated in New Zeland, then moved to Europe, and his first
research paper on logic was published at the age already
thirty-eight years.)}
~\cite{Prior-1996}, p.~46). Further, one can
continue reading: ``and leaves it to existentialists and to
comparatively informal 'dialecticians' to study the more 'dynamic'
aspect of reality''. So it seems that it is encouraging enough to
include temporal logic in the consideration of preference models.
After building a preference model in temporal logic, one can
analyze it using a deductive approach. The goal is searching, if
any, for contradictions in a model or analyzing its satisfiability.
It is also possible to
inference about correctness of preference objectives. Thus,
temporal logic creates new possibilities for analysis of
preferences by going beyond the static world of classical logic.
It also allows to illustrate the dynamic aspect of preferences
which describe situations of preference valuations that vary over
flows of time.

The issue of preference models based on temporal logic are discussed in some
works~\cite{Klimek-2013-peccs,Klimek-Wojnicki-Ernst-2013-icaisc,Klimek-2013-icmmi}.
For example, in work~\cite{Klimek-2013-peccs} some basic notions and definitions
are introduced.
The architecture of an inference system is proposed.
It allows examining of satisfiability, contradiction or
being a tautology of preference models.
The methodology for gathering information about preferences
in the requirements engineering process is proposed in work~\cite{Klimek-2013-icmmi}.
The quality of the requirements engineering process strongly influences
in the positive way the quality of the entire developed system.
Finally,
in work~\cite{Klimek-Wojnicki-Ernst-2013-icaisc},
it is shown that preference modeling could reduce the state space of the agent-based world.

Let us present briefly temporal logic which constitutes
important theoretical background of the approach presented in the work.
\emph{Temporal Logic} TL which is a branch of symbolic logic focusses on
statements whose valuations depend on time flows,
e.g.~\cite{Wolter-Wooldridge-2011},
which has strong application in the area of software engineering.
It is used for the system analysis where behaviors of events are of interest.
TL exists in many varieties, however, considerations in this paper are
limited to the \emph{Linear Temporal Logic} LTL,
i.e.\ logic with the linear time structure.

The syntax of LTL logic is formulated over a countable set of
\emph{atomic formulas} $AP=\{ p, q, r, ...\}$ and
the set of \emph{temporal operators} ${\cal M}=\{ \som, \alw \}$.
\begin{definition}
\label{def:LTL-syntax}
A \emph{LTL formula} is a formula which is build using the following rules:
\begin{itemize}
\item if $p \in AP$ then $p$ is a LTL formula,
\item if $p$ and $q$ are formulas, then $\neg p$, $p \dis q$,
      $p \con q$, $p \imp q$, $p \equ q$ are LTL formulas,
\item if $p$ is a formula, then $\circledast\, p$, where $\circledast \in {\cal M}$, is also a LTL formula.
\end{itemize}
\end{definition}
Thus, the whole \emph{LTL alphabet} consists of the following symbols:
$AP$, ${\cal M}$ and classical logic symbols like $\neg$, $\dis$, $\con$, etc.
There is no difficulty to introduce other symbols, e.g.\ parenthesis,
which are omitted here to simplify the presentation.
The ${\cal M}$ set consists two fundamental and unary temporal logic operators,
where $\som$ means ``sometime (or eventually) in the future''
and $\alw$ means ``always in the future''.
Considerations in the work are focused on the LTL logic,
and particulary on \emph{Propositional Linear Temporal Logic} PLTL.

The semantics of the LTL logic is traditionally defined using
the concept of \emph{Kripke structure}
which is considered as a graph, or path, whose nodes represent the reachable states
$w=s_{0},s_{1},s_{2},...$,
or in other words the reachable worlds,
and a labeling function which maps each node to
a set of atomic formulas $2^{AP}$ that are satisfied in a state.
A \emph{valuation} function $\nu(w(i)) \longrightarrow 2^{AP}$, where $i\geq 0$,
and $w(i)$ means the $i$-th element of the path $w$,
allows to define the \emph{satisfaction} $\models$ relation between
a path and a LTL formula, e.g.\
$w\models p$ iff $p\in w(0)$,
$w\models \neg p$ iff it is not $p\in w(0)$ and
$w\models \som p$ iff $p\in w(i)$, where $i\geq 0$, etc.
Theorems and laws of the LTL logic can be found in many works, e.g.~\cite{Emerson-1990}.
Considerations in this paper are limited
to the \emph{smallest\emph{, or \emph{minimal},} temporal logic}, e.g.~\cite{Chellas-1980}.

Logic and reasoning are cognitive skills.
Logical reasoning is the process of using
a sound mathematical procedures to given statements to arrive at conclusions.
There are some techniques, or proof procedures, which are systematic
methods producing proofs in some calculus, or provable, statements.
Although the work is not based on any particular method of reasoning,
the method of semantic tableaux is presented in a more detailed way.
The method of \emph{semantic tableaux}, or \emph{truth tree}
is well known in classical logic but it can be applied
in modal logic~\cite{Agostino-etal-1999}.
The method is based on the formula decomposition
using predefined decomposition rules.
At each step of the well-defined procedure,
formulas become simpler as logical connectives are removed.
At the end of the decomposition procedure,
all branches of the received tree are searched for contradictions.
When the branch of the truth tree contains a contradiction,
it means that the branch is \emph{closed}.
When the branch of the truth tree does not contain a contradiction,
it means that the branch is \emph{open}.
When all branches are closed, it means that the tree is closed.
Simple examples of inference trees are shown in Fig.~\ref{fig:truth-trees}.
The adopted decomposition procedure,
as well as labelling, refers to the first-order predicate calculus
and can be found in work~\cite{Hahnle-1998}.

The semantic tableaux method can be treated as a \emph{decision procedure},
i.e.\ the algorithm that can produce the polar answer Yes-No as a response to some important questions.
Let $F$ be an examined formula and ${\cal T}$ is a truth tree build for a formula.
\begin{corollary}{}
\label{th:decision-procedures}
The semantic tableaux method gives answers to the following questions related to the satisfiability problem:
\begin{itemize}
\item formula $F$ is not satisfied iff the finished ${\cal T}(F)$ is closed;
\item formula $F$ is satisfiable iff the finished ${\cal T}(F)$ is open;
\item formula $F$ is always  valid iff finished ${\cal T}(\neg F)$ is closed.
\end{itemize}
\end{corollary}
The semantic tableaux method is based on the systematic search for models that satisfy a formula.
To show that a formula is unsatisfiable, it needs to
show that all branches are closed.
Hence, if the tree is closed, it means there is not model that satisfy a formula.
To show that a formula is satisfiable, it needs to find one open branch.
If the tree is open, it means there exist a model that satisfy a formula.
If the tree for the negation of a formula is closed,
it means there is no model that satisfy a formula,
and as a result of the fact that this is a proving by contradiction,
it leads to the conclusion that the initial formula is always valid.

\section{Context-awareness of the IoT}
\label{sec:context-IoT}

A proposal to construct preference models that operate in the IoT basing on
temporal logic, graph structures and the multi-agent cooperation is introduced in this Section.
Preference models are expressed in terms of temporal logic formulas and can be
changed on demand during the system operation.

\begin{figure}[htb]
\centering
\begin{pspicture}(9.0,7.0) 
\scalebox{0.8}{
\rput(0.0,0.0){\psframe[linecolor=gray,fillcolor=gray,fillstyle=solid](1.5,1.5)}
\rput(10.1,0.5){\psframe[linecolor=gray,fillcolor=gray,fillstyle=solid](0.9,6.0)}
\rput(1.0,5.0){\psframe[linecolor=gray,fillcolor=gray,fillstyle=solid](5.0,2.0)}
\rput(5.0,0.0){\psframe[linecolor=gray,fillcolor=gray,fillstyle=solid](1.0,3.0)}
\rput(2.0,2.0){\psframe[linecolor=gray,fillcolor=gray,fillstyle=solid](1.5,1.5)}
\rput(7.0,5.0){\psframe[linecolor=gray,fillcolor=gray,fillstyle=solid](2.0,1.0)}
\rput(6.0,0.0){\psframe[linecolor=gray,fillcolor=gray,fillstyle=solid](1.0,0.4)}
\psline(0,0)(3,0)
\psline(4,0)(7,0)
\psline(8,0)(11,0)
\psline(11,0)(11,0.5)
\psline(11,6.5)(11,7)
\psline(6,7)(11,7)
\psline(0,7)(1,7)
\psline(0,5)(0,7)
\psline(0,4)(0,0)

\rput(8.0,0.5){\psframe(1.0,0.45)}
\rput(8.0,1.0){\psframe(1.0,0.45)}
\rput(8.0,1.5){\psframe(1.0,0.45)}
\rput(8.0,2.0){\psframe(1.0,0.45)}
\rput(8.0,2.5){\psframe(1.0,0.45)}

\rput(6.1,0.5){\psframe(0.9,0.45)}
\rput(6.1,1.0){\psframe(0.9,0.45)}
\rput(6.1,1.5){\psframe(0.9,0.45)}
\rput(6.1,2.0){\psframe(0.9,0.45)}
\rput(6.1,2.5){\psframe(0.9,0.45)}
\rput(8.0,3.5){\psframe(1.0,0.4)}
\rput(8.0,4.0){\psframe(1.0,0.4)}
\rput(8.0,4.5){\psframe(1.0,0.4)}\rput(8.5,4.7){\small p018}
\rput(6.1,3.5){\psframe(0.9,0.4)}
\rput(6.1,4.0){\psframe(0.9,0.4)}
\rput(6.1,4.5){\psframe(0.9,0.4)}
\rput(5.0,3.5){\psframe(1.0,0.4)}
\rput(5.0,4.0){\psframe(1.0,0.4)}
\rput(5.0,4.5){\psframe(1.0,0.4)}\rput(5.5,4.7){\small p015}

\rput(9.1,0.5){\psframe(0.9,0.45)}
\rput(9.1,1.0){\psframe(0.9,0.45)}
\rput(9.1,1.5){\psframe(0.9,0.45)}

\rput(4.0,0.5){\psframe(0.9,0.45)}
\rput(4.0,1.0){\psframe(0.9,0.45)}\rput(4.5,2.2){\small p010}
\rput(4.0,1.5){\psframe(0.9,0.45)}
\rput(4.0,2.0){\psframe(0.9,0.45)}
\rput(4.0,2.5){\psframe(0.9,0.45)}

\rput(2.0,0.5){\psframe(1.0,0.45)}
\rput(2.0,1.0){\psframe(1.0,0.45)}
\rput(2.0,1.5){\psframe(1.0,0.45)}

\rput(1.0,2.0){\psframe(0.9,0.45)}
\rput(1.0,2.5){\psframe(0.9,0.45)}
\rput(1.0,3.0){\psframe(0.9,0.45)}

\rput(1.0,3.6){\psframe(0.9,0.45)}
\rput(2.0,3.6){\psframe(0.9,0.45)}

\rput(7.5,0.0){g1}
\rput(3.5,0.0){g2}
\rput(0.0,4.5){g3}
}
\end{pspicture}
\caption{A sample parking space}
\label{fig:graph-parking-space}
\end{figure}
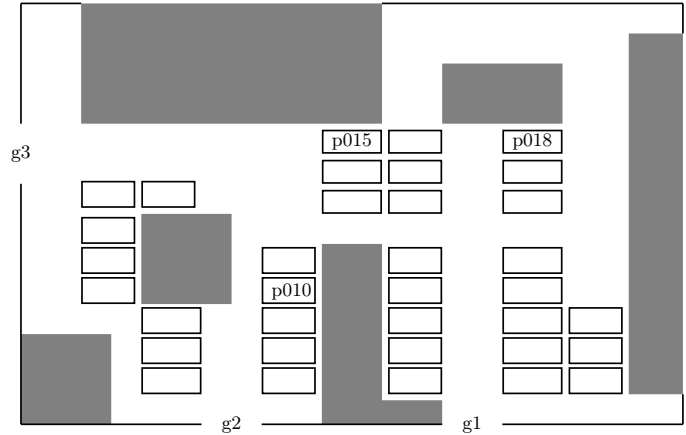
\begin{figure}[htb]
\centering
\begin{pspicture}(7.5,2.5) \showgrid
\psset{linecolor=blue,fillcolor=blue,fillstyle=solid}

\rput(0.0,0.0){\psframe[linecolor=gray,fillcolor=gray,fillstyle=none](8,3)}
\rput(.8,2.0){\rnode{r1}{}}
\rput(1.9,2.7){\rnode{r2}{}}
\ncline[linewidth=1pt,linecolor=white]{-}{r1}{r2}\naput[labelsep=-13pt,nrot=:U]{\textsf{Graph layer}}
\rput(6.6,2.6){\rnode{r3}{}}
\rput(7.3,2.1){\rnode{r4}{}}
\ncline[linewidth=1pt,linecolor=white]{-}{r3}{r4}\naput[labelsep=-13pt,nrot=:U]{\textsf{Graph layer}}

\rput(1.0,0.5){\psframe(.4,.6)}
\rput(1.5,0.5){\psframe(.4,.6)}
\rput(2.0,0.5){\psframe(.4,.6)}
\rput(2.5,0.5){\psframe(.4,.6)}
\rput(3.0,0.5){\psframe(.4,.6)}
\rput(3.5,0.5){\psframe(.4,.6)}
\rput(4.2,.8){\textsc A1}
\rput(4.5,0.5){\psframe(.4,.6)}
\rput(5.0,0.5){\psframe(.4,.6)}
\rput(5.5,0.5){\psframe(.4,.6)}
\rput(6.0,0.5){\psframe(.4,.6)}
\rput(6.5,0.5){\psframe(.4,.6)}
\rput(7.0,0.5){\psframe(.4,.6)}

\rput(1.8,1.2){\psframe(.6,.6)}
\rput(3.0,1.2){\psframe(.6,.6)}
\rput(4.8,1.2){\psframe(.6,.6)}
\rput(6.0,1.2){\psframe(.6,.6)}
\rput(4.2,1.5){\textsc A2}

\rput(2.8,2.0){\psframe(3,.6)}
\rput(4.2,2.3){\textcolor{white}{\textsc A3}}
\end{pspicture}
\caption{Agents and their hierarchy operating in a graph structure}
\label{fig:agent-model}
\end{figure}
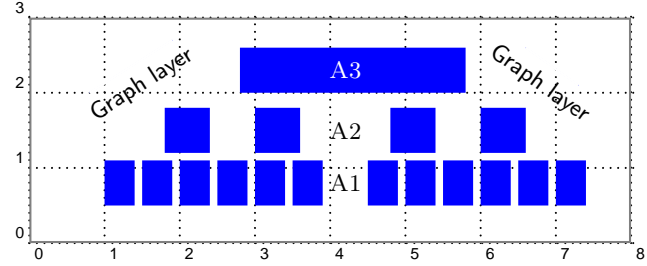

Let us consider a sample parking space,
c.f.\ Fig.~\ref{fig:graph-parking-space}.
It consists of some entrance/exit gates,
and a number of identified parking areas.
The outline of the procedure used for the management of preference models could be described as
a sequence of some steps that are performed concurrently,
and are briefly discussed below.
The world of things/objects is modeled using a graph structure,
the definition of which is given in this Section.
The procedure is based on the following ideas:
\begin{itemize}
\item
The system records basic events of a smart parking space,
and events are recorded in particular nodes of a graph structure;
\item
Events are registered in nodes (of the IoT) and the time-stamp for
every event is also registered.
Identification of users/inhabitants is based on various possibilities,
such as RFID, PDA devices, biometric data, image scanning and pattern recognition, and others.
The issue of user/inhabitant identification is not discussed here;
\item
The event information is collected permanently and allow to build logical specications
expressed in terms of temporal logic formulas based on the corresponding algorithm.
Specification provides knowledge about the user behavior
and allows to prepare the preference decision for a newly observed user
that appear in the smart parking space.
\item
When a gate of the parking space is reached by a new user/inhabitant,
then the logical specification is analyzed to prepare a preference decision,
and pass this decision to the user/inhabitant.
Generating preference decision is made on the highest level of the agent hierarchy,
c.f.\ Fig.~\ref{fig:agent-model}.
Generating decisions take into account the knowledge about the past behaviour,
which is logically analyzed using, for example, the semantic tableaux method,
which allows to search for satisfiability (choice a preference decision based on past behaviours),
or contradictions (new behaviors overriding the previous behaviors).
\end{itemize}


The goal mentioned above, we suggest to implement by the following multiagent system.
The world of things/objects is modeled using a graph structure,
that glues the cooperation of three types of agents into one
system. Fig.~\ref{fig:agent-model} shows the whole agent world,
i.e.\ agents that operate in the smart environment.
It is assumed the existence of the following types of agents:
\begin{description}
\item[$A3$] -- agent also called \emph{decision agent},
      it exist permanently in the system and its primary aim is to
      prepare/compute preference-based decisions for
      a new user/inhabitant entering the parking space,
      these decisions are based on the gathered knowledge expressed
      in terms of graph representation structure and logical specifications which are prepared by agents $A2$,
      decision agents can also modify knowledge, what is their secondary aim, when they find that
      the newly observed behaviors include contradictions in regards to the old behaviors,
      i.e.\ knowledge expressed in (old) logical formulas,
      and the contradiction elimination might be a result of the formal analysis of logic formulas
      using, for example, the semantic tableaux method.
      Each of these agents marks some graph nodes and any graph transformation that use such a node cooperate with the pointed agent.
\item[$A2$] -- agents also called \emph{follower agents},
      they might exist temporary in the system and their aim is to
      observe objects that appear in the smart environments and
      built logical specifications considered
      as a set of temporal logic formulas that express behaviours of newly observed users/inhabitants,
      the logical specification constitutes knowledge about
      user preferences and is built based on information form agents $A1$. They are generated when some event occurs.
\item[$A1$] -- agents also called \emph{reactive agents}, or \emph{node agents},
      they exist permanently in the system and their aim is to
      operate in an individual node  gathering information about users/inhabitants
      which reach this node in the IoT's network,
      information is obtained through sensors and combined with the identification of a user/inhabitant.
      The determination by such agent of some event causes the execution of some graph transformation.
\end{description}

The graph layer is defined as a labelled and attributed graph (abbrev.\ LA-graph) defined below.
\begin{definition}
\label{An-LA-graph-is}
An \emph{LA-graph} is a labelled and attributed
digraph of the following form $G=(V,E,\{lab_{X},att_{X}\}_{X=V,E})$, such that:
\begin{itemize}
  \item $V$ is a finite and nonempty set of vertices;
  \item $E\subset V\times V$ is a set of directed edges (arcs);
  \item $lab_{X}:X\rightarrow\mathcal{L}_{X}$ are labelling functions for
nodes ($X=V)$ and edges ($X=E)$ respectively, where $\mathcal{L}_{V},\mathcal{L}_{E}$
are sets of node and edge labels;
  \item $att{}_{X}:X\rightarrow2^{\mathcal{A}_{X}}$ are attributing functions
for nodes ($X=V)$ and edges ($X=E)$ respectively, where $\mathcal{A}_{V},\mathcal{A}_{E}$
are sets of node and edge attributes.
\end{itemize}
\end{definition}
The interpretation of labels and attributes in Definition~\ref{An-LA-graph-is}
is following. A label $l\in\mathcal{L}$ identifies unambiguously
a given vertex/edge, e.g., by assigning an unique name to an object,
an attribute $a\in\mathcal{A}$ is some property of a vertex/edge,
for example a power station transformer output voltage or time interval
of a smart meter. As stated in Definition~\ref{An-LA-graph-is}, one
may assign a set of attributes to a given entity. It should be stressed
that an attribute $a$ must not be confused with its value. Thus the
notion of LA-graph may be compared to a class definition. The graph
analog of a class instance is an instantiated LA-graph defined
below.
\begin{definition}
Let $G=(V,E,\{lab_{X},att_{X}\}_{X=V,E})$ be an LA-graph.
An \emph{instantiation} of G is a triple $\hat{G}=(G,val_{V},val_{E})$, where $val_{X}:X\times\mathcal{A}_{X}\rightarrow\Omega_{X}$
is an instantiating function for nodes ($X=V)$ and edges ($X=E)$
respectively.
$\hat{G}$ will be also referred to as an \emph{instantiated LA-graph} (shortly, ILA-graph).
\end{definition}

The mentioned idea we will explain on the idea of the parking system.
the graph consist of only four types of nodes:
\begin {itemize}
\item node labelled by G -- that describe a gateway to the parking,
\item node labelled by R -- that describe an road segment,
\item node labelled by P -- that describe a parking place,
\item node labelled by C -- that describe a car.
\end {itemize}

In the real solutions we have to considered also a few types od sensors, the area of their cooperation,
but it will influence only to more complex behaviour of the agent of $A1$ type (so we will not consider them here).
We assume that with each node labelled by $G$, $R$ or $P$ are associated agents of type $A1$ that discover
the appearance of a car in the space described by it.
The structure of the system is maintained inside the graph maintained by the $A3$.
The more complex action is associated with appearance of a car in
the gateway (coming for outside of the parking); it consist from the sequence of actions:
\begin{itemize}
    \item a new node labelled by $C$ is added to a graph -- it is linked with node labeled by $G$,
    \item a new agent of a type $A2$ is created, and it communicates with the agent of type $A3$
    supervising this gateway -- asking for the preference of the identified car.
    If the preferred place is empty -- this information we can find from the graph structure
    (i.e.\ nor node labeled by $C$ is linked with this parking place) -- than the information is
    passed to the car, and it start to take a parking place.
    If the preferred place is occupied,
    that this agent try to find the solution the closed to this preference.
    This generated agents follows by the car, observing the driver behaviour
    both while it goes to the parking place and while it leaves the parking.
    \item when the car leaves the parking, agent of the $A2$ type sent its observation to
    agent of $A3$ type, and destroy itself.
\end{itemize}
The actions associated with the activities of agent of $A1$ (associated with nodes labeled by $R$ and $P$) is simpler, agent notice movement of the car and:
\begin{itemize}
    \item perform a graph transformation changing edge linking a node representing a car
    with the previous detected position to the edge linking the considered node with the node representing a car,
    \item informs agent of $A2$ type following the car about its behaviour.
\end{itemize}
We assume that each agent gathers some part of knowledge from other agents.
Due to the large size of the maintained graph, representing the knowledge,
in the real application it will be maintained by a set of agents cooperating over
the replicated complementary graph structure.
Formally it takes or replicates some part of the graph specifying this knowledge.
Next agents modify its local representation in a parallel way.
This leads a problem with synchronization of their work in the distributed system.
Fortunately at least one solution Replicated Complementary Graph~\cite{Kotulski-Sedziwy-2011-icannga}
allows to split the graph representation onto
a few subgraphs, remembering border and replicated nodes in such a way that:
\begin{itemize}
\item it is possible gluing these subgraphs to the centralized form,
\item explicit synchronization of the transformation rules applied in the parallel way,
\item extend the current knowledge either by the replication of some information.
\end{itemize}

Let us consider a simple yet illustrative example for the approach.
Let us present rules for the $A1$\marginpar{$A1$} agents,
i.e.\ the node agents.
These agents occupy the lowest level in the whole hierarchy of the agent activities.
Agents are assigned to particular nodes of parking space/graph structure.
The basic events that refer to the presence of users/inhabitants are recorded in nodes.
Let $O=\{ o_{1},o_{2},... \}$ is a set of users/inhabitants identified
in the system.
Individual users have unique identifiers.
The problem of unambiguous identification is well known and is not be discussed widely here.
Let $D= \{ d_{1}, d_{2}, ... \}$ is a set of events,
where every $d_{i}$ belongs to $\langle O,V,T \rangle$,
where $O$ is a set of identified users/inhabitants,
$V$ is a node of a network,
and $T$ is a set of time stamps.
For example,
$d_{i}=\langle idKR55,p0018,t2014.01.28.09.30.15 \rangle$
means that the presence of the $idKR55$ object is observed at
the physical point/area $p0018$ of the parking space,
and the time stamp assigned to this event is $t2014.01.28.09.30.15$.
Summing up,
the primary task of every agent $A1$ is event logging in a particular node of a parking space.

Let us present rules for the $A2$\marginpar{$A2$} agents,
i.e.\ the follower agents.
These agents occupy the middle level in the whole hierarchy of the agent activities.
Agents gather knowledge about preferences of users/inhabitants in the considered area.
Preferences are expressed in terms of temporal logic formulas.
To obtain such logical specifications, the information produced by agents $A1$,
i.e.\ events registered in particular nodes,
are processed.
Informally speaking,
the $A2$ agents translate physical events to logical specifications.
The input for this translation are events $d_{i}$ as defined above.
The output are logical formulas understood as triples of the form
$l_{i}=\langle id,f,r \rangle$,
where $id$ is an identifier/id of an object that operates in the parking space/IoT,
$f$ is a temporal logic formula, i.e.\ PLTL, and
$r$ is a number of occurrences of this formula as a result of a user behaviour.
The algorithms for generating logical specification are to be a subject
of future works, c.f.\ also concluding remarks in the last Section.
The entire logical specification is a set of these triples,
i.e.\ $S=\{ l_{i} : i \geq 0 \}$.
The introduced notion requires some explanation.
Many users can be identified in the entire system.
The system stores information about different users and
the $id$ allows to differentiate formulas
intended for a particular user.
The meaning of $f$ is obvious,
i.e.\ it is a syntactically-correct temporal logic formula.
The $r$ element, where $r>0$, is a kind of counter and
it means multiple occurrences of a given formula as a result
of multiple registrations of the same behaviour.
For example,
$\langle idKR55,g2 \imp\som p018,7 \rangle$ and
$\langle idKR55,g2 \imp\som p015,2 \rangle$
means that user $idKR55$ enters gate $g2$ and
sometime reaches the parking area $p018$ (seven times in the past), and
sometime reaches the parking area $p015$ (two times in the past).
When the preference decision is taken,
and if $p018$ is free,
then this parking area is suggested as the most preferred one,
otherwise first $p015$ or second no suggestion is made.

Let us present rules for the $A3$\marginpar{$A3$} agent,
i.e.\ the decision agent.
This agent occupy the highest level in the whole hierarchy of the agent activities,
and its purpose is to prepare preference decisions for a user/inhabitant.
The agent analyze knowledge about preferences expressed in terms of logical formulas,
which are produced by agents $A2$.
The input for this analysis is a logical specification which is produced by
the follower agents.
The output are preference decisions prepared for a user/inhabitant.


\begin{figure}[htb]
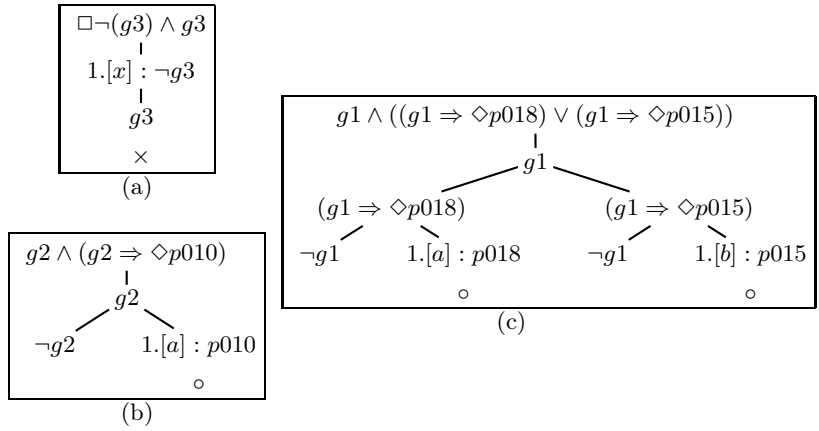

\begin{tabular}{ll}
\begin{minipage}{.3\linewidth}
\centering
\framebox{
\pstree[levelsep=4.5ex,nodesep=2pt,treesep=25pt]
       {\TR{$\alw\neg (g3) \con g3$}}
       {\pstree{\TR{$1.[x]: \neg g3$}}
       {\pstree{\TR{$g3$}}{$\times$}}}}\\
       (a)\\\mbox{}\\
\framebox{
\pstree[levelsep=4.5ex,nodesep=2pt,treesep=25pt]
       {\TR{$g2 \con (g2 \imp \som p010)$}}
       {\pstree{\TR{$g2$}}
         {{\TR{$\neg g2$}}{\pstree{\TR{$1.[a]: p010$}}{$\circ$}}}}}\\
         (b)
\end{minipage}
&
\begin{minipage}{.5\linewidth}
\centering
\framebox{
\pstree[levelsep=4.5ex,nodesep=2pt,treesep=25pt]
       {\TR{$g1 \con ((g1 \imp\som p018) \dis (g1 \imp\som p015))$}}
       {\pstree{\TR{$g1$}}
         {{\pstree{\TR{$(g1 \imp\som p018)$}}{{\TR{$\neg g1$}}{\pstree{\TR{$1.[a]:p018$}}{$\circ$}}}}
         {\pstree{\TR{$(g1 \imp\som p015)$}}{{\TR{$\neg g1$}}{\pstree{\TR{$1.[b]:p015$}}{$\circ$}}}}}}}\\
         (c)
\end{minipage}
\end{tabular}
\caption{Some sample truth trees}
\label{fig:truth-trees}
\end{figure}
Let us consider some cases to explain some ideas of the Algorithm.
Say, the logical specification for a user $o_{i}$ contains
logical formula $\alw\neg(g3)$ which means that the user never entered gate $g3$.
However,
when at a certain time point user $o_{i}$ appears at $g3$,
then it provides the logical formula $\alw\neg (g3) \con g3$ which might give
the reasoning tree for the semantic tableaux method shown in
Fig.~\ref{fig:truth-trees}.a, c.f.\ closed branch ($\times$).
Of course,
this tree could be a part of a greater truth tree,
which is omitted here to simplify considerations,
but it must contain at least one closed branch,
i.e.\ branch that contains a contradiction.
It follows that the logical specification should be modified
removing formula $\alw\neg (g3)$ from the initial specification,
then new formula which results of a new event,
entering gate $g3$, are to be added to specification.
Another case could refer a situation when user enters gate $g2$
and the logical specification contains formula $g2 \imp\som p010$,
which means that when $g2$ is reached then sometime area $p010$ is reached.
It leads to the following formulas and reasoning:
$g2 \con (g2 \imp\som p010) \Longrightarrow \som p010$,
or using the truth tree Fig.~\ref{fig:truth-trees}.b,
c.f.\ the open branch ($\circ$).
The preference decision is the sample $p010$ parking area, if it is still free.
The last case is the situation when a gate is reached
and there exist two (or more) different (sub-)formulas,
i.e.\ $g1 \con ((g1 \imp\som p018) \dis (g1 \imp\som p015)) \Longrightarrow \som p018 \dis \som p015$,
or using the truth tree Fig.~\ref{fig:truth-trees}.c,
c.f.\ the open branches ($\circ$).
It means that both $p018$ and $p015$ are areas of preference.
It also means that the last element of the triple $l_{i}$ which is frequency $r$ of
a particular formula determines which parking area is chosen as a preferred one,
if it is still free.

\section{Conclusions}

In the paper we show an example of application of the IoT concept in
the multi-agent environment where the external knowledge is represented by
a graph and the preference model of behaviours is represented by the Linear Temporal Logic.
Despite its exploratory nature, this work offers some formal foundations for future work.
Although the current study is based on a relatively small sample,
the findings suggest that it can be deployed in the case of more complex ones.

Temporal logic enables representing behaviours of objects in terms of temporal information within a logical framework.
This allows to avoid many ambiguities typical for natural languages.
This also allows to perform the reasoning process in a formal and reliable way.
Logic modeling involves the development and application of formal logics to
solve problem of interest in a field of the IoT.
Temporal logic has strong application in the area of software engineering,
and is used for the system analysis where behaviors of events are of interest.
In work~\cite{Dwyer-etal-1999} the method of discovering behaviours considered as patterns is discussed.
It provides logical specifications expressed in terms of temporal logic.
In work~\cite{Finger-dovGabbay-1992} the temporalisation aspect of a plain system is discussed.
It can be also used to analyze normal system events and behavior mining in smart environments.
Both these works might provide strong support for algorithms generating logical specifications
for users/inhabitants' behaviours in the IoT network, considered as a context-aware and pro-active system.

The graph representing the parking consists of a few dozen of nodes,
but in real systems such a graph representation consist of hundreds of
thousands nodes.
Thus, such a graph should be divided onto smaller parts that will
cooperate themselves (with the explicit synchronization mechanism).
Such an environment called Replicated Complementary graphs is supported by
GRADIS~\cite{Kotulski-Sedziwy-2009-icdcs} multiagent framework where each agent controls one local graph $G_{i}$.
Following the FIPA~\cite{FIPA-2014} specification~\cite{FIPA-Specifications-2014} we assume
a very simple functionality of a multi-agent environment,
reduced to a message transport and a broker system.
This approach is similar to those applied in popular frameworks like JADE~\cite{JADE-2014} or
Retsina~\cite{Retsina-2014}.
Summing up,
it seems legitimate the statement that
the deployment of these two formalisms,
i.e.\ temporal logic as a branch of formal logic and graph representation and transformations,
provide the synergistic effect,
i.e.\ acting together the total effect is greater than if taken separately.

\subsubsection*{Conflict of interests.}
The authors declare that there is no conflict of interests regarding
the publication of this article.

%
\label{sect:bib}
\bibliographystyle{plain}
\bibliography{../bib/rk-bib-rk,../bib/rk-bib-main,../bib/rk-bib-pervasive,../bib/rk-bib-graph}

\end{document}